\documentclass{ws-procs9x6}

\makeatletter
\def\bfm#1{\mbox{\boldmath $#1$}}

\makeatother

\begin{document}
\title{  \vspace*{-0.5cm}
  \hfill{\normalsize\vbox{%
    \hbox{\rm\small YITP-04-09}
  }}\\
  \vspace{0.2cm}Role of Strange Quark Mass\\
	in Pairing Phenomena in QCD%
\footnote{\uppercase{T}alk given at \uppercase{KIAS-APCTP}
	\uppercase{I}nternatinal \uppercase{S}ymposium on
	\uppercase{A}stro-\uppercase{H}adron 
	\uppercase{P}hysics
 	``{\it \uppercase{C}ompact \uppercase{S}tars:~%
	\uppercase{Q}uest \uppercase{F}or \uppercase{N}ew
	\uppercase{S}tates of \uppercase{D}ense \uppercase{M}atter}\,'', 
	\uppercase{N}ovember 10-14, 2003, 
 	\uppercase{KIAS}, \uppercase{S}eoul, \uppercase{K}orea.
	\uppercase{T}his Talk is based on 
	\uppercase{R}ef.~1.
	}}

\author{Hiroaki Abuki}

\address{Yukawa Institute for Theoretical Physics,\\
Kyoto University, Kyoto 606-8502, Japan\\ 
E-mail: abuki@yukawa.kyoto-u.ac.jp}

\maketitle

\newcommand{\beq}{\begin{eqnarray}}
\newcommand{\eeq}{\end{eqnarray}}
\newcommand{\e}{\epsilon}
\newcommand{\ds}[1]{
  \setbox0=\hbox{\ensuremath{#1}}
  \hbox to\wd0{\hbox to0pt{\hbox to\wd0{\hss/\hss}\hss}\box0}}
\newcommand{\isum}{\int\hspace{-14.0pt}\sum} 
\newcommand{\isums}{\int\hspace{-10.5pt}\sum} 
\newcommand{\bra}[1]{\left\langle\left. #1 \right\vert\right.}
\newcommand{\ket}[1]{\left.\left\vert #1 \right. \right\rangle}
\newcommand{\braket}[1]{\left\langle #1\right\rangle}

\abstracts{
We study the dynamical effect of strange quark mass as well as 
 kinematical one on the color-flavor unlocking transition using 
 a NJL model.
Paying a special attention to the multiplicity of gap parameters,
 we derive an exact formula of the effective potential for 5-gap 
 parameters.
Based on this, we discuss that the unlocking transition might be 
 of second order rather than of first order as is predicted by a
 simple kinematical criterion for the unlocking.}

\section{Introduction}
The dynamics which quark-gluon matter exhibits under high baryon 
  density is one of the most challenging and exciting problems
  in QCD \cite{BL84,Iwasaki,Reviews}.
A number of literatures have revealed the realization of the
  color-flavor locked (CFL) type of the pairing order \cite{CFL} 
  for sufficiently large quark number chemical potential
  \cite{Reviews,Shuster,Schafer,Hsu1,Hsu2,Hong}.
In contrast to this solid fact, however, the phases for large, 
  and realistic value of $M_s^2/\mu$ are still veiled in mystery, 
  and so many phases and the associated phase transitions have 
  been studied so far
  \cite{unlock_SW,unlock_AR,Kcond,LOFF,g2SC,gCFL,Matsuura}.
The unlocking transition from the CFL phase to the 2SC state is
  simplest one of the examples.
This transition with increasing the value of $M_s^2/\mu$ is 
  firstly investigated using Nambu--Jona-Lasinio (NJL) type effective
  models \cite{unlock_SW,unlock_AR}.
Their results show that the 1st order unlocking transition takes place
  at some critical $M_s^c$, and a simple kinematical picture for the
  critical mass works in quite satisfactory way.
This criterion is based on the conjecture that the transition occurs 
  when the mismatch of the Fermi momenta of light and strange quark 
  becomes as large as the magnitude of the gap.
In Ref.~\refcite{unlock_SW}, the expansion in flavor breaking parameter 
 $M_s^2/\mu$ is used to the construction of the fermionic dispersions
 and the effective potential. 
On the other hand, in Ref.~\refcite{unlock_AR}, the coupled gap equation
  is exactly solved with non-perturbative treatment of $M_s$.
However, their construction of the effective potential is quite 
  ambiguous and, as we shall show, possibly fails because of the 
  multiplicity of gap parameters.
Even though their results are in good accordance with each other, 
  there still exist a possibility that the true solution is missed.

In this talk, we revisit the unlocking transition in more complete way
  than others \cite{unlock_SW,unlock_AR}, by making a proper use of 
  the Pauli construction of the effective potential.
In particular, we study how the CFL state and other states are 
  realized in the multi-gap parameter space and how the potential 
  gets distorted with increasing the value of $M_s^2/\mu$.
More complete analyses have been made in Ref.~\refcite{HabukiNJ}.

\section{Coupled Gap Equation and Effective Potential}\label{sec:2}
In this section, we present an outline for obtaining the gap equations
  and the effective potential with which we can determine the ground state
  of the system characterized by $(\mu,T,M_s)$ parameter.

\noindent
{\bf\sf Self-energies.~}
We first introduce the color-flavor mixed quark base:~%
$
  q^\alpha=\sum_{a,i}{\big(\lambda_\alpha\big/{\sqrt{\mathstrut 2}}\big)^i_a}q^a_i,
$
where $\lambda_\alpha (\alpha=1\sim 8)$ are the Gell-Mann matrices
and we defined $(\lambda^9)^a_i=\sqrt{\frac{2}{3}}\delta^a_i$.
Then the Nambu-Gor'kov propagator in the 2-compnent quark field
$Q=(q,\bar{q}^t)$ is written as
\beq
  \bfm{S}(q_0,\bfm{q})&\equiv&i\,{\rm F.T.}\langle
  T\{Q(x)\bar{Q}(0)\}\rangle=\left(
  \begin{matrix}
    \ds{q}+\ds{\mu}-{M} & \Sigma(q_0,\bfm{q})\cr
    \gamma_0\Sigma(q_0,\bfm{q})^\dagger\gamma_0 & \ds{q}^t-\ds{\mu}^t+{M}\cr
  \end{matrix}\right)^{-1}.\label{eq:NGP}
\eeq
$\Sigma(q_0,\bfm{q})$ is off-diagonal self-energy which gives rise to
gaps in the quasi-quark dispersion. 
$M$ is the quark mass matrix in the color-flavor mixed base 
$M_{\alpha\beta}={\rm
tr}\left[\lambda_\alpha\hat{m}\lambda_\beta\right]/2$,
where $\hat{m}={\rm diag.}(0,0,M_s)$ is mass matrix in flavor space.
The pure CFL ansatz for the off-diagonal self-energy is expressed
in the color-flavor mixed base as following, 
\beq
  \Sigma^{\rm CFL}_{\alpha\beta}(q_0,\bfm{q})%
     =C\gamma_5\otimes\left(\begin{matrix}
         \Delta_8{\bf 1}_8 &  \cr
          &  \Delta_1\cr
       \end{matrix}\right).\label{eq:CFLansatz}
\eeq
Here, $C$ is the charge conjugation matrix $C=i\gamma_2\gamma_0$.
$C\gamma_5$ guarantees that the pairing takes place in the 
  $J^P=0^+$ channel.
On the other hand, we would have the 2SC state with $u$-$d$ pairing 
  $\Sigma^{{\rm 2SC}_{ud}}=-\epsilon_{ij}\epsilon^{ab3}\Delta_{\rm
  2SC}=\Delta_A\epsilon_{ij}\epsilon^{ab3}$
  for sufficiently large value of $M_s$, which is written as
\beq
   \Sigma^{{\rm 2SC}_{ud}}_{\alpha\beta}(q_0,\bfm{q})%
     =C\gamma_5\otimes\left(\begin{matrix}
          \Delta_{\rm 2SC}{\bf 1}_3 &  &  & \cr
          & {\bf 0}_4   &  &\cr
	  &  &  -\frac{1}{3}\Delta_{\rm 2SC} & -\frac{\sqrt{2}}{3}\Delta_{\rm 2SC}\cr
	  &  &  -\frac{\sqrt{2}}{3}\Delta_{\rm 2SC} & -\frac{2}{3}\Delta_{\rm 2SC}\cr
       \end{matrix}\right),\label{eq:2SCansatz}
\eeq
in the color-flavor mixed base.
From the two expressions for the 2SC and CFL phases, it is quite natural
  to expect the distorted CFL state ($d$CFL), the minimal interpolating
  pairing ansatz between those phases, for the small but finite strange
  quark mass \cite{unlock_SW,unlock_AR},
\beq
  \Sigma^{d\rm CFL}_{\alpha\beta}(q_0,\bfm{q})%
     \stackrel{M_s<M_s^c}{=}%
       C\gamma_5\otimes\left(\begin{matrix}
     \Delta_{83}{\bf 1}_3&  &    &  \cr
     &  \Delta_{82}{\bf 1}_4&    &  \cr
     &  &   \Delta_{81}& \Delta \cr
     &  &    \Delta &  \Delta_{11}\cr
     \end{matrix}\right).\label{eq:ANSATZ}
\eeq
In this parameterization, the $SU(3)_{C+V}$ symmetric CFL state and
  the 2SC phase with $SU(2)_C\times SU(2)_L\times SU(2)_R$ symmetry 
  are expressed only as different limits of the $d$CFL phase which is 
  invariant under color flavor simultaneous rotation $SU(2)_{C+V}$.
By introducing the 5-dimensional vector $
   \vec{\Delta}=(\Delta_{83},
                \Delta_{82},
		\Delta_{81},
		\Delta,
		\Delta_{11})^t$, 
we can express the model space for the pure CFL state as
  $\vec{\Delta}_{\rm
  CFL}=(\Delta_{8},\Delta_{8},\Delta_{8},0,\Delta_{1})^t$, 
  which spans a 2-dimensional planer section,
while the 2SC phase as the 1-parameter vector 
  $\vec{\Delta}_{\rm 2SC}=(\Delta_{\rm 2SC},0,-\Delta_{\rm
  2SC}/3,-\sqrt{\mathstrut 2}\Delta_{\rm 2SC}/3,-2\Delta_{\rm 2SC}/3)^t$.

\vspace*{0.2cm}
\noindent
{\bf\sf Gap equations.~}
The anomalous propagator is defined by the
off-diagonal element of the Nambu-Gor'kov propagator,
which takes the following form for our ansatz Eq.~(\ref{eq:ANSATZ}),
\beq
  S_{12}^{\alpha\beta}(q)%
  =i{\rm F.T.}\langle q^\alpha(x)q^\beta(0)\rangle
  =\gamma_5 C\otimes\left(\begin{matrix}
                    R_{83}{\bf 1}_3& &  &\cr
		     & R_{82}{\bf 1}_4&  &\cr
		     & & R_{81} & R \cr
		     & & R & R_{11} \cr
                     \end{matrix}\right).\label{eq:FPO}
\eeq
$\{R_{83},R_{82},R_{81},R,R_{11}\}$ are propagatores, which are
  complicated functions of $(q,\vec{\Delta},\mu,M_s)$ \cite{HabukiNJ}. 
The gap equation is obtained by using Feynman rule for the self
  consistency between proposed self-energy and the one-loop self-energy.
In the case of NJL model with interaction vertex for one-gluon exchange,
  ${\mathcal L}_{\rm int}=(1/2)g^2\bar{q}\gamma_\mu t^a q\bar{q}\gamma^\mu
  t^a q$, we obtain
\beq
  \Sigma_{\alpha\beta}=i4g^2\int\frac{d^4q}{(2\pi)^4}%
    (T^a)_{\gamma\alpha}S^{\gamma\delta}_{12}(q)(T^a)_{\delta\beta}.
\eeq
Here, the bare vertex in the color-flavor mixed base is defined by
$(T^a)_{\alpha\beta}={\rm
  tr}\left[\lambda_\alpha\lambda_a\lambda_\beta\right]/4$.
This integral matrix equation contains the following set of five
  equations.
\beq
   \Delta_{83}&=&g^2%
      \frac{1}{12}\left(5\bfm{\mathcal R}_{83}-\bfm{\mathcal R}_{81}%
      -2\sqrt{2\mathstrut}\bfm{\mathcal R}-2\bfm{\mathcal
      R}_{11}\right)\equiv g^2K_{83}[\vec{\Delta}],\label{eq:NGEQ1}\\
   \Delta_{82}&=&g^2\frac{1}{12}%
      \left(2\bfm{\mathcal R}_{82}+2\bfm{\mathcal R}_{81}%
      +\sqrt{2\mathstrut}\bfm{\mathcal R}-2\bfm{\mathcal R}_{11}\right)%
      \equiv g^2K_{82}[\vec{\Delta}],\label{eq:NGEQ2}\\
   \Delta_{81}&=&g^2\frac{1}{12}%
      \left(-3\bfm{\mathcal R}_{83}+8\bfm{\mathcal R}_{82}%
      -\bfm{\mathcal R}_{81}+2\sqrt{2\mathstrut}\bfm{\mathcal R}%
      -2\bfm{\mathcal R}_{11}\right)\equiv
      g^2K_{81}[\vec{\Delta}],\label{eq:NGEQ3}\\
   \Delta&=&g^2\frac{\sqrt{2\mathstrut}}{12}%
          \left(-3\bfm{\mathcal R}_{83}+2\bfm{\mathcal R}_{82}%
	  +\bfm{\mathcal R}_{81}-\sqrt{2\mathstrut}\bfm{\mathcal
	  R}\right)\equiv g^2K[\vec{\Delta}],\label{eq:NGEQ4}\\
   \Delta_{11}&=&g^2\frac{1}{6}\Big(-3\bfm{\mathcal
          R}_{83}-4\bfm{\mathcal R}_{82}-\bfm{\mathcal
	  R}_{81}\Big)\equiv g^2K_{11}[\vec{\Delta}].\label{eq:NGEQ5}
\eeq
Here, we have defined $\bfm{\mathcal R}$ by
\beq
   \bfm{\mathcal R}_i(\vec{\Delta};M_s)%
      =i\int\frac{d^4q}{(2\pi)^4}\frac{1}{4}{\rm tr}%
      \left[C\gamma_5R_i(q;\vec{\Delta},M_s)\right]. \label{eq:R}
\eeq

\noindent
{\bf\sf Effective potential for multi-gap parameters.~}
Here, we do not attempt to formulate the Pauli-construction of the
  effective potential, which is exactly done in Ref.~\refcite{HabukiNJ}. 
But instead, we illustrate how we would miss the true effective 
  potential due to the multiplicity of gap parameters when it is
  constructed by the integration of the gap equation, 
  and only show the correct procedure to obtain true one.

We might think that the derivative of the effective potential is
\beq
   \frac{\partial\Omega_{\rm eff}}{\partial\Delta_i}\propto(\mbox{gap
   equation for $\Delta_i$}).
\eeq
If so, the effective potential is constructed by
\beq
   \Omega_{\rm eff}\big[\vec{\Delta}\big]%
   \propto\sum_i\frac{\Delta_i^2}{2g^2}-\sum_i%
   \int_0^1 dt\Delta_iK_i[t\vec{\Delta}].
\eeq
But this gives a false formula in the case that the many gap parameters
exist and couple each other by different couplings.

Actually, the derivative of the effective potential coincides with 
some linear combination of the gap equation like
\beq
   g^2D^{-1}_{ij}\frac{\partial\Omega_{\rm
   eff}}{\partial\Delta_j}=A_{ij}(\Delta_j-g^2K_j[\vec{\Delta}]),
\eeq
where the diagonal matrix $D={\rm diag.}(3,4,1,2,1)$ represents
  the degeneracies of the gap parameters,
and the matrix $A$ is relating the gap 
parameters
$\vec{\Delta}=(\Delta_{83},\Delta_{82},\Delta_{81},\Delta,\Delta_{11})$
and the corresponding condensates
$\vec{\phi}=(\phi_{83},\phi_{82},\phi_{81},\phi,\phi_{11})$ as
$4g^2\vec{\phi}=\hat{A}\vec{\Delta}$.
For the detail of the definition $\vec{\phi}$ and the form of the matrix
$A$, see Ref.~\refcite{HabukiNJ}.
We can obtain the true formula for the effective potential as
\beq
  \Omega_{\rm
  eff}[\vec{\Delta}]&=&\sum_{i,j}\frac{(DA)_{ij}}{2g^2}\Delta_i\Delta_j%
  -\sum_{ij}\int_0^1 dt (DA)_{ij}\Delta_iK_j[t\vec{\Delta}].\label{eq:eff}
\eeq
$DA$ is real-valued symmetric matrix, whose eigenvalues are all real.

\section{Numerical Results}\label{sec:nr}
Here, we present our numerical results.
The cutoff parameter is set to $\Lambda=800$MeV, and the coupling
  parameter $g^2$ is tuned to reproduce $400$MeV for the constituent
  mass of quark at zero chemical potential \cite{unlock_AR}.

\subsection{Solution of gap equation}\label{ssec:sge}
\begin{figure}[tp]
 \begin{minipage}{0.49\textwidth}
  \includegraphics[width=0.99\textwidth,clip]{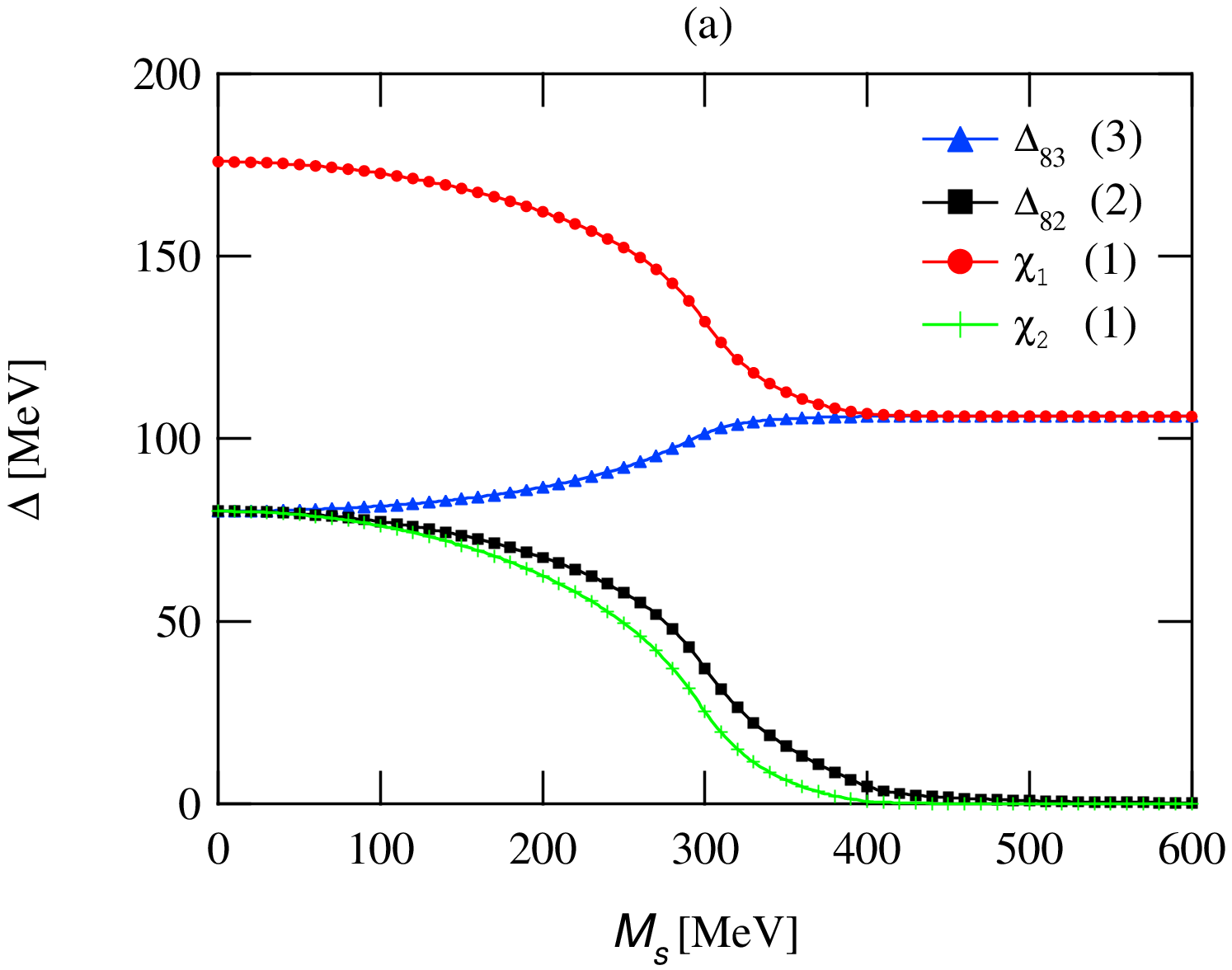}
 \end{minipage}%
 \hfill%
 \begin{minipage}{0.49\textwidth}
  \includegraphics[width=0.99\textwidth,clip]{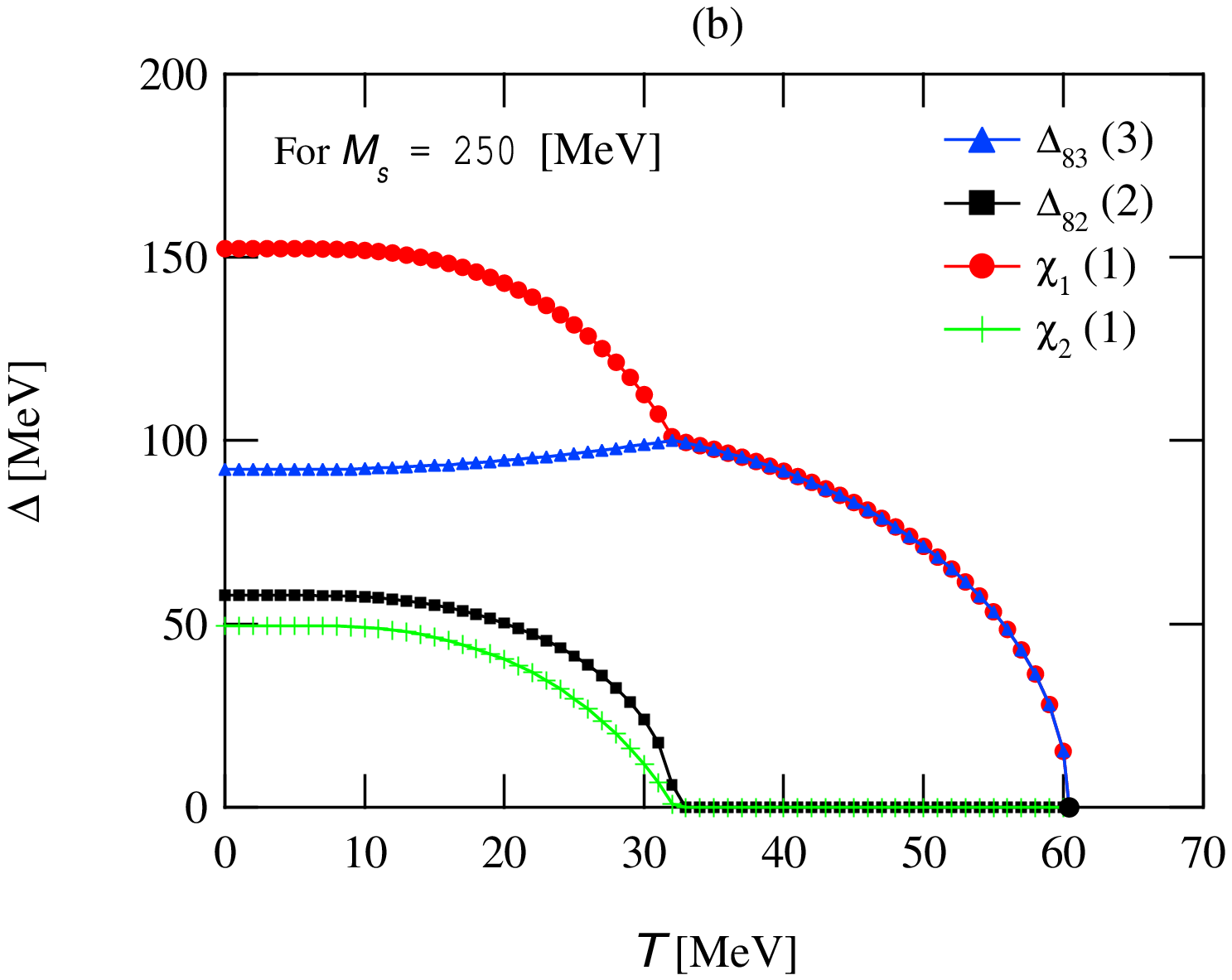}
 \end{minipage}
 \caption[]{(a) Gaps are plotted as functions of $M_s$.
 (b) The temperature dependence of the gaps for $M_s=250$
 MeV. }
 \label{fig:02}
\end{figure}

We display the solution of the coupled gap equation
  Eqs.~(\ref{eq:NGEQ1})$\sim$(\ref{eq:NGEQ5}).
In FIG.~\ref{fig:02}(a), we show the eigenvalues of the original gap
  matrix Eq.~(\ref{eq:ANSATZ}) as functions of $M_s$.
$(\Delta_{83}, \Delta_{82})$ are the gaps for iso-triplet and doublet modes.
$(\chi_1, \chi_2)$ are the eigenvalues of the iso-singlet mixing sector.
According to the values of these parameters, the states are distinguished by
\begin{center}
\begin{tabular}{c||c||c}
  Gap parameters&Global Symmetry&States\\ \hline
  $\Delta_{83}=\Delta_{82}=\chi_2,\,\chi_1\neq 0$&$SU(3)_V$&CFL for $M_s=0$\\
  $\Delta_{83}=\chi_1,\,\Delta_{82}=\chi_2=0$&$SU(2)_L\times SU(2)_R$&2SC for $M_s>M_s^c$\\
  otherwise &$SU(2)_V$&distorted CFL ($d$CFL)
\end{tabular}
\end{center}

\vspace*{0.2cm}
\noindent
{\bf\sf Continuous Color-Flavor Unlocking.~}
It is surprising that the system seems to undergo continuous transition
from the CFL to the 2SC, and actually stays in the distorted CFL
($d$CFL) phase even for $M_s>\mu=400$ MeV.
We will see later that even in our case, the 2SC is always a solution of
  the gap equation, but with higher energy than the CFL state.
In FIG.~\ref{fig:02}(b), we display in FIG.~\ref{fig:02}, the gaps
  $(\Delta_{83},\Delta_{82},\chi_1,\chi_2)$ for $M_s=250$ MeV as 
  functions of temperature.
We see that as the temperature is raised, the $d$CFL phase encounters
  the continuous transition to the 2SC state at $T=35$MeV, and
  eventually the system undergoes 2nd order phase transition to the
  normal fermi-liquid at $T=60$MeV.

\subsection{Effective potential for multi-gap parameter space}%
\begin{figure}[tp]
 \begin{minipage}{0.4\textwidth}
  \includegraphics[width=0.95\textwidth,clip]{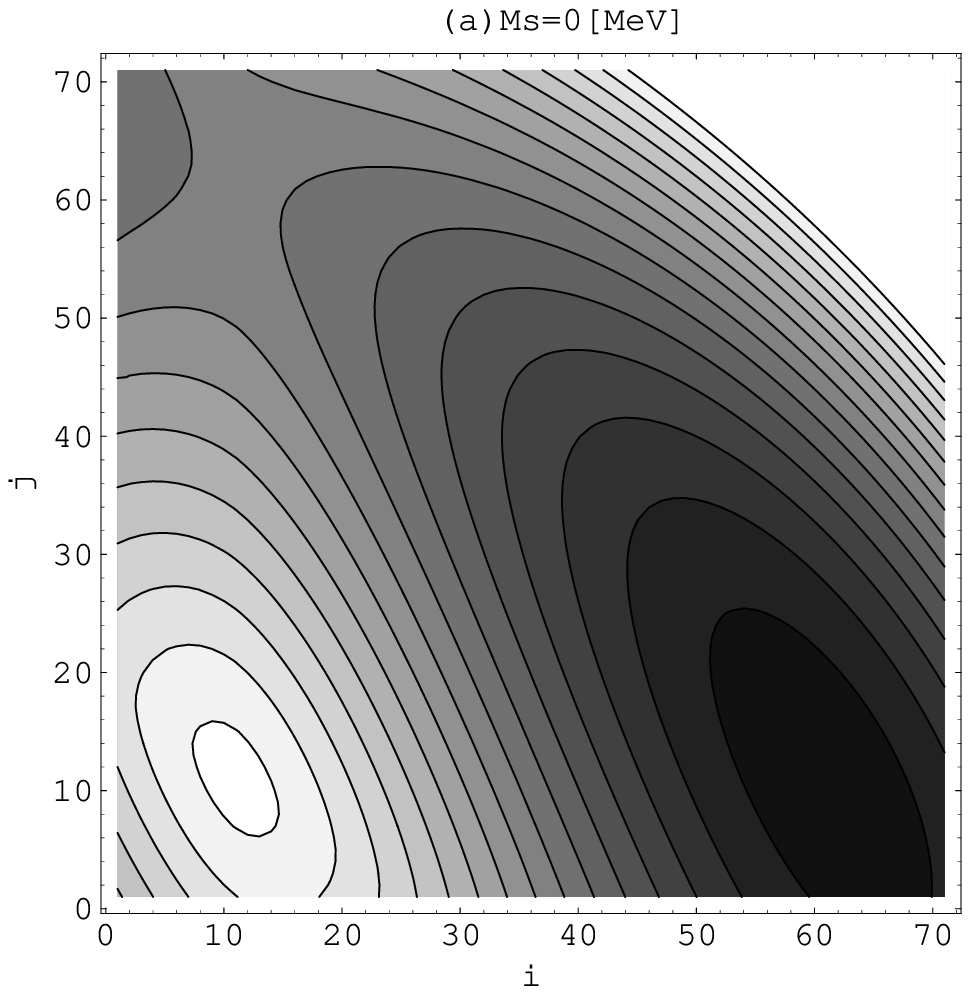}
 \end{minipage}%
 \hfill%
 \begin{minipage}{0.4\textwidth}
  \includegraphics[width=0.95\textwidth,clip]{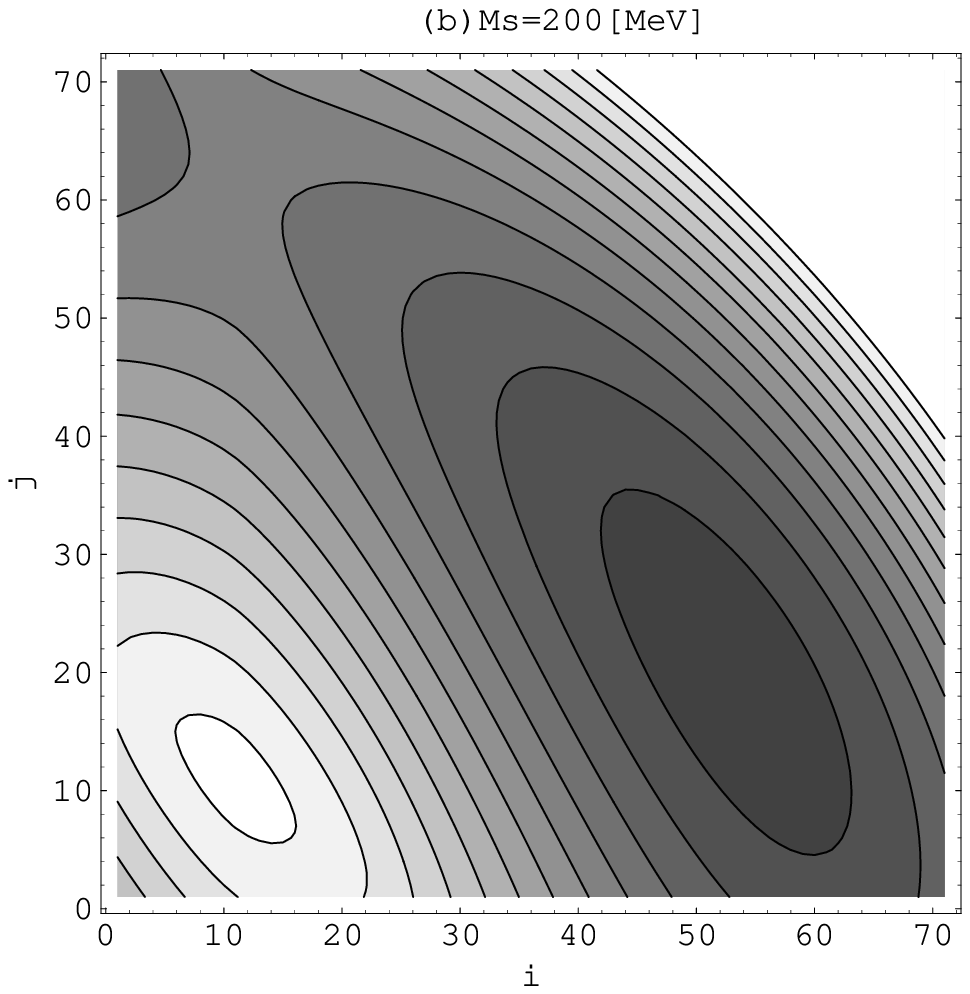}
 \end{minipage}
 \hfill%
 \begin{minipage}{0.19\textwidth}
  \includegraphics[width=0.8\textwidth,clip]{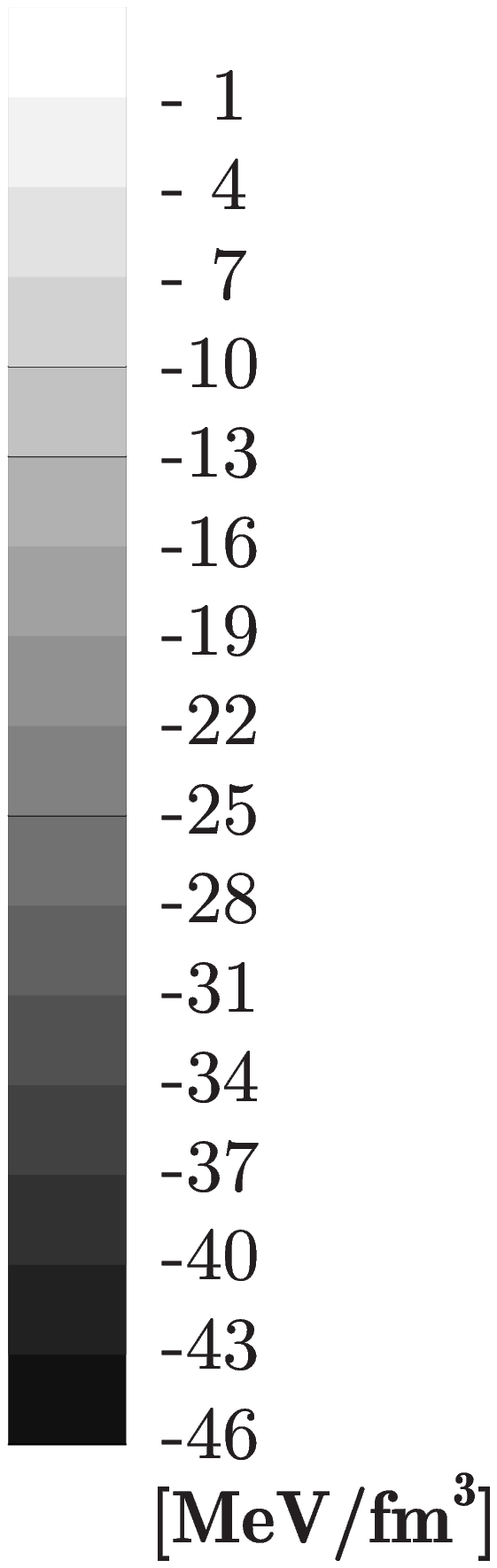}
 \end{minipage}
 \begin{minipage}{0.4\textwidth}
  \includegraphics[width=0.95\textwidth,clip]{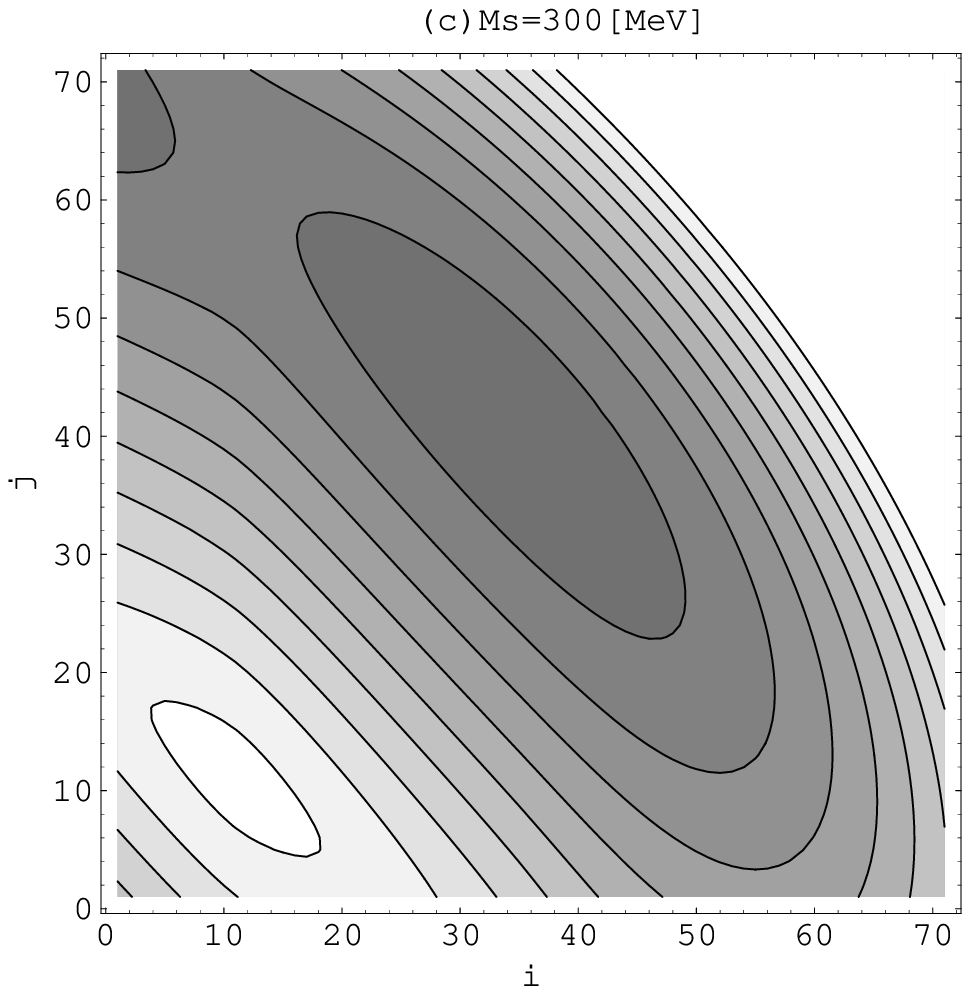}
 \end{minipage}%
 \hfill%
 \begin{minipage}{0.4\textwidth}
  \includegraphics[width=0.95\textwidth,clip]{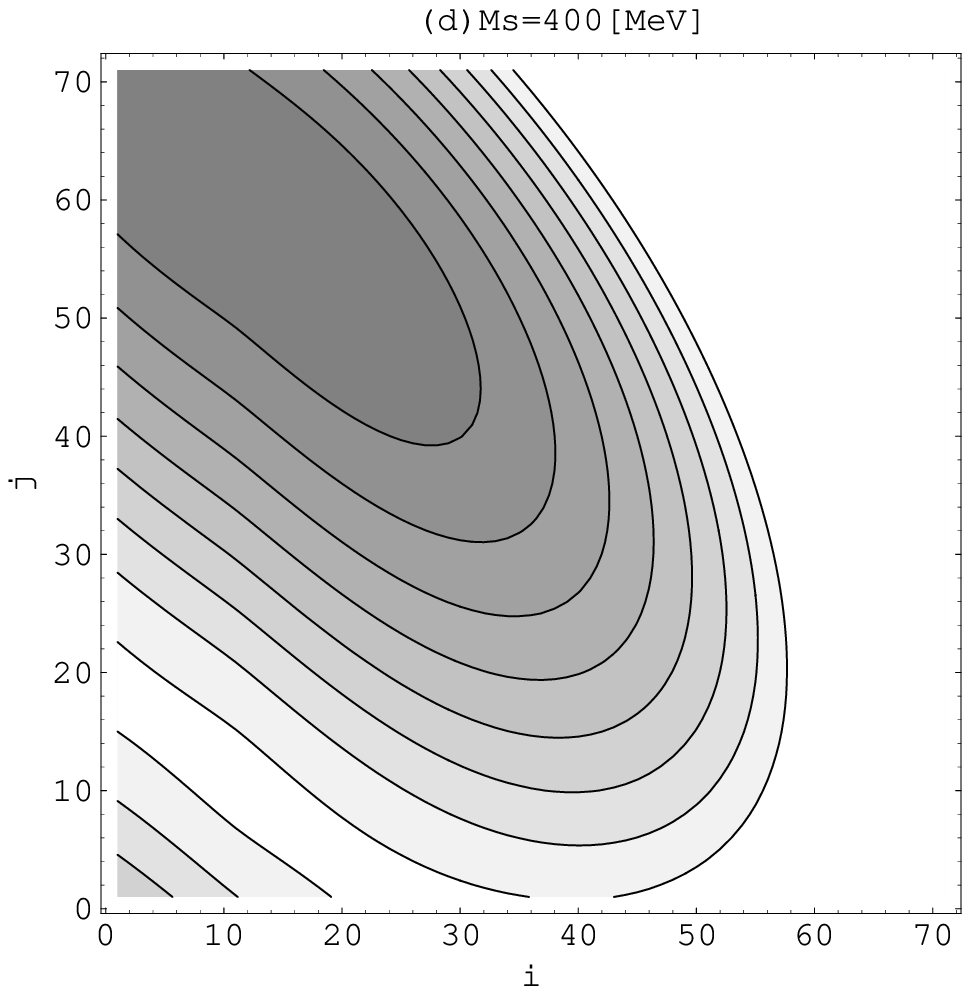}
 \end{minipage}
 \hfill%
 \begin{minipage}{0.19\textwidth}
  \includegraphics[width=0.8\textwidth,clip]{indicator.eps}
 \end{minipage}
 \caption[]{Contour plot of the effective potential 
 $\Omega_{\rm eff}[\Delta(i,j)]$ where $\Delta(i,j)$ is defined by
 Eq.~(\ref{eq:section}), for $M_s=0$ MeV (a), $M_s=200$ MeV (b), $M_s=300$
 MeV (c), and for $M_s=400$ MeV (d). 
 }
 \label{fig:04}
\end{figure}

\label{sec:Unstable}
We now study how these states are realized in the
  multi-gap parameter space by computing the effective potential 
  with help of the formula Eq.~(\ref{eq:eff}).
Let us first introduce the vector
\beq
   \vec{\Delta}(i,j)=\frac{(i-10)\vec{\Delta}_{\rm
   CFL}+(j-10)\vec{\Delta}_{\rm 2SC}}{50},\label{eq:section}
\eeq
with the CFL gap vector $\vec{\Delta}_{\rm CFL}=(80,80,80,80,-175)$
  and the 2SC gap vector $\vec{\Delta}_{\rm 2SC}=(106,0,-35,-50,-70)$
  for $Ms=0$ (in the chiral limit).
Eq.~(\ref{eq:section}) define 2-dimensional planer section in the
  5-dimensional gap parameter space, which includes the simple Fermi gas
  $(10,10)$ with $\vec{\Delta}=\vec{\bf 0}$, the CFL $(60,10)$ and the
  2SC $(10,60)$.
In FIG.~\ref{fig:04}, we show the contour plot of the effective
  potential $\Omega_{\rm eff}[\vec{\Delta}(i,j);M_s]$,
  for $M_s=0$ MeV (a),~$200$ MeV (b),~$300$ MeV (c) and $400$ MeV (d).
We can see that the CFL state $(60,10)$ looks the global minimum in this
  2-parameter space $(i,j)$ for $M_s=0$ MeV.
On the other hand, the 2SC state $(10,60)$ is realized as a saddle point
  which is stable in the direction of the simple Fermi gas $(10,10)$,
  but is unstable in the direction towards the CFL state $(60,10)$.
As the strange quark mass is increased to $200$ MeV, the CFL minimum
  moves towards the 2SC point, and its condensation energy gets reduced,
  while the position and the energy of the 2SC state is unaffected by
  $M_s$.
This minimum point $(55,20)$ is expected to be located close to the
  distorted CFL $d$CFL state in the full 5-parameter space.
The $d$CFL state gets distorted significantly towards the 2SC state
  at $M_s=300$ MeV (FIG.~\ref{fig:04}(c)), and seems absorbed into
  the 2SC as the strange quark mass approaches the order of the
  chemical potential $\sim\mu$ of the system.
We can conclude that the unlocking transition is not of 1st order.
More detailed analyses made in Ref.~\refcite{HabukiNJ} reveals that
  this transition is of 2nd order and indicates that the
  $d$CFL satate at low strange quark density is the Bose-Einstein
  condensation of tightly bound pairs rather than the BCS state 
  \cite{AbukiHatsuda}.

\vspace*{0.2cm}
\noindent
{\bf\sf 2SC as a solution of gap equation.~}
What should be stressed here is that the 2SC stays at least a saddle
  point in the full model space irrespective of the value of $M_s$.
This means that the 2SC is always a solution of the gap equation, and
  if we had missed in obtaining the correct effective potential, then we
  misunderstood it as the true ground state instead of the $d$CFL state
  for intermediate strange quark mass $M_s$.

\begin{figure}[tp]
 \begin{minipage}{0.49\textwidth}
  \includegraphics[width=0.99\textwidth,clip]{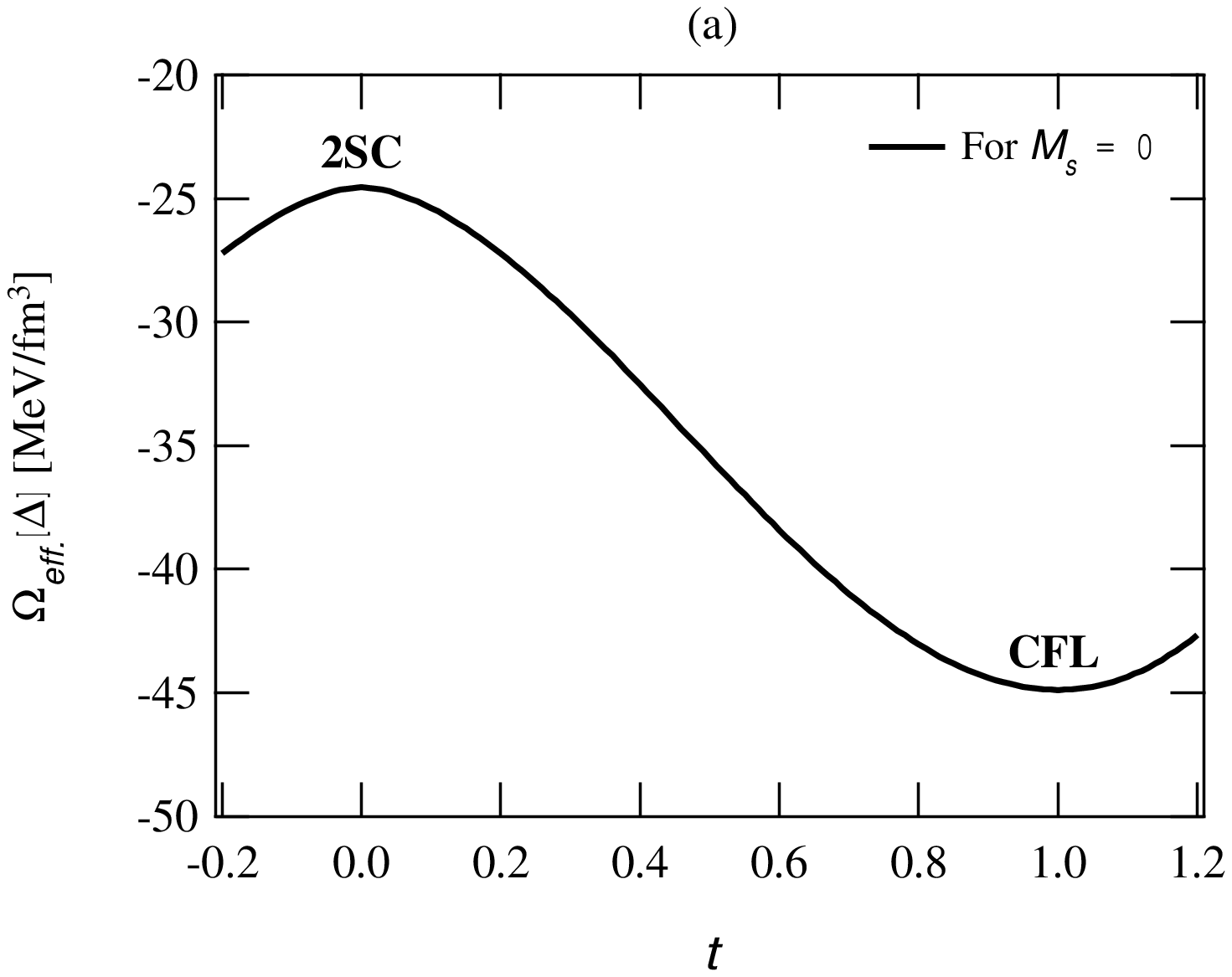}
 \end{minipage}%
 \hfill%
 \begin{minipage}{0.49\textwidth}
  \includegraphics[width=0.99\textwidth,clip]{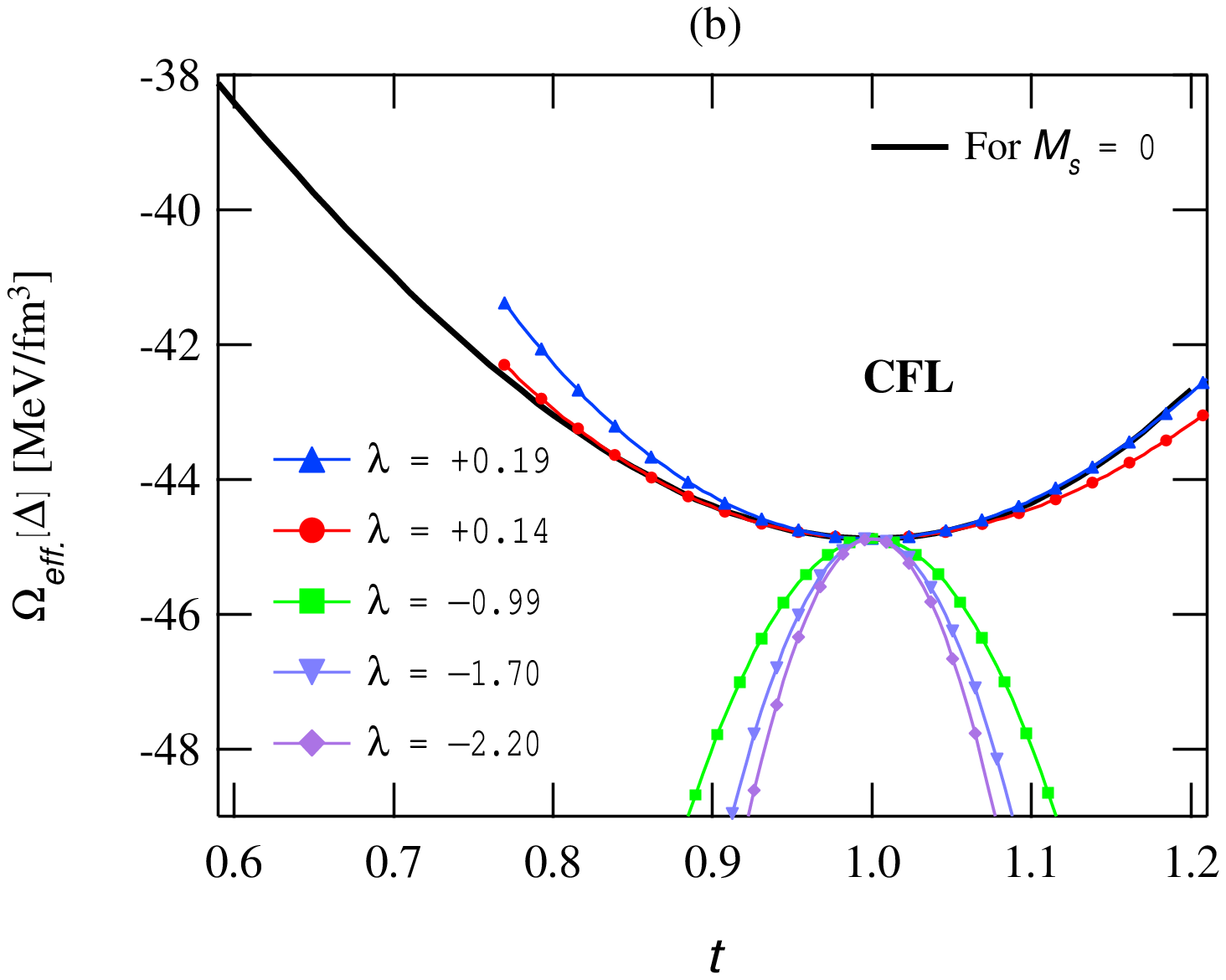}
 \end{minipage}
 \caption[]{(a) The values of the effective potential on the
 1-dimensional line connecting from the 2SC state $t=0$ to the CFL state
 $t=1$. (b) The bold line is just an enlargement of the potential curve
 for $M_s=0$ in (a), and five set of the points (triangles, bold dots,
 squares, reflected triangles, lozenges) represent the potential
 curves in the directions of the eigenvectors belonging to the five
 eigenvalues $\lambda\cong(0.19,\,0.14,\,-0.99,\,-1.70,\,-2.20)\mu^2$ 
 of the Hessian curvature matrix at the CFL.
 }
 \label{fig:05}
\end{figure}

\vspace*{0.2cm}
\noindent
{\bf\sf Unstable CFL state?~}
In FIG.~\ref{fig:05}(a), we plot the section of the effective potential
  surface FIG.~\ref{fig:04}(a) on the line linking the 2SC $(10,60)$
  $\vec{\Delta}_{t=0}=\vec{\Delta}_{\rm 2SC}$ and the CFL $(60,10)$
  $\vec{\Delta}_{t=1}=\vec{\Delta}_{\rm CFL}$. 
Both states are determined by solving the gap equation in 5-gap
  parameter space for $M_s=0$ MeV.
Now we address the question whether the CFL state is true global minimum
  even in the full model space or not.
FIG.~\ref{fig:05}(b) shows us the answer for this question.
Five potential curves are shown near the pure CFL for $M_s=0$,
each of which represents the potential curve in the direction of an
  eigenvector belonging to the corresponding eigenvalue $\lambda$ of the
  Hessian curvature matrix at the CFL point $
  \frac{\partial^2\Omega}{\partial\Delta_i\partial\Delta_j}\Big|%
  _{\vec{\Delta}=\vec{\Delta}_{\rm CFL}}.$
What is surprising is that the CFL state is not the global minimum in
  the full model space.
It is unstable in three out of five directions.
These three directions correspond to the color-flavor sextet channels
  in which the interaction acts repulsive \cite{HabukiNJ}.
Also it should be noted that the color-flavor sextet gap parameters
  do not break any symmetry, thus are not the order parameter for the
  unlocking as is anticipated in their critical behaviour 
  $\sim (1-T/T_c)^{3/2}$ near critical temperature $T_c$ in the chiral
  limit \cite{HabukiSD}.
Anyway, because the effective potential becomes unbound once the sextet
  condensations are included, it makes no physical sense to include
  symmetric mean fields contribution \cite{CFL,Shov,Gatto} in the ansatz
  for the gap matrix.
It is said that the sextet condensations are not induced within the mean
  field approximation, whereas they might be triggered by higher order
  fluctuation effects beyond the mean fields.

\section{Conclusion}\label{sec:summary}
We have adopted the NJL model to study the phases of quark matter under
  high chemical potential.
By making proper use of the Pauli-construction method, we have derived
  the exact formula for the effective potential for multi-gap parameters.
In particular, we have studied the unlocking phase transition from 
  the CFL to the 2SC.
We list main results below.

\vspace*{0.2cm}
\noindent
{\bf\sf 1.~Second order phase transition.~} 
The unlocking transition is of continuous weak 2nd order. 
	  The 1st order unlocking never appear even at $T=0$.
	  This contradicts the simple kinematical picture for the
	  color-flavor unlocking transition.

\vspace*{0.2cm}
\noindent
{\bf\sf 2.~Toughness of the CFL state.~} 
The CFL state at $T=0$ is much more robust against the increase
	  of $M_s$ than is predicted from the kinematical criterion, 
	  and the 2SC state is continuously connected from the $d$CFL
	  state at the strange quark mass $M_s^c>\mu\,(\gg 
	  2\sqrt{\mu\Delta_8(\mu,M_s=0)})$. 

\vspace*{0.2cm}
\noindent
{\bf\sf 3.~The 2SC as a saddle point.~}
2SC state is always a saddle point, a solution of the gap
	  equation, of the effective potential for any value of $M_s$,
	  which is unstable in the direction towards the $d$CFL state,
	  as well as three color-flavor sextet directions. 
	  The $d$CFL state, a solution of the gap equation with a larger
	  condensation energy than the 2SC state approaches the 2SC
	  saddle point as $M_s$ approaches $M_s^c$. 

\vspace*{0.2cm}
\noindent
{\bf\sf 4.~Role of the sextet gap parameters.~}
The $d$CFL state is also unstable in the directions for three
	  color-flavor symmetric channels.
	  This is attributed to the fact that the interaction is
	  repulsive in these channels. We should not include the symmetric
	  components of the gap parameter into our ansatz from the
	  beginning, because no Cooper instability is present in these
	  channels due to the absence of the attractive force.
	  However, the effects of the sextet gaps on the anti-triplet
	  sector are very small even if they are included.

\vspace*{0.2cm}
\noindent
In this talk, we completely neglected the electric charge neutrality
  and also the color neutrality \cite{CN}.
It would be interesting to include these effect into the gap equation,
  and to study how our $d$CFL phase is robust or fragile to be
  withdrawn by the neutrality condition.
Also, we have ignored the usual chiral condensates in the vacuum.
However, we have to include this effect to discuss the phase transitions
  in the lower chemical potential region.
Studying phase transitions in this regime by improving our model along
  this line is also an interesting subject to be done in the future.

\section*{Acknowledgments}
I am grateful to T.~Kunihiro for stimulating discussions.
I would also like to thank Deog-Ki Hong, Sung-Ah Cho and the
  other organizers and assistants for giving
  me an opportunity to give a talk at International Symposium on
  Astro-Hadron Physics 
  and also for their hospitality during my stay at Seoul.


\begin{thebibliography}{0}
\bibitem{HabukiNJ}H.~Abuki, arXiv:hep-ph/0401245.
\bibitem{BL84} D.~Bailin and A.~Love, Phys. Rep. {\bf 107}, 325 (1984),
\bibitem{Iwasaki}
	M.~Iwasaki and T.~Iwado, Phys. Lett. B {\bf 350}, 163 (1995);
	M.G.~Alford, K.~Rajagopal and F.~Wilczek, Phys. Lett. B {\bf 422},
	247 (1998);
	R.~Rapp, T.~Sch\"afer, E.V.~Shuryak and M. Velkovsky,
	Phys. Rev. Lett. {\bf 81}, 53 (1998). 
\bibitem{Reviews} For reviews, see K.~Rajagopal and F.~Wilczek,
	arXiv:hep-ph/0011333; M.G.~Alford,
	Ann. Rev. Nucl. Part. Sci. {\bf 51}, 131 (2001);
	T.~Sch\"afer, arXiv:hep-ph/0304281; D.H.~Rischke,
	arXiv:nucl-th/0305030.
\bibitem{CFL}
	M.G.~Alford, K.~Rajagopal and F.~Wilczek, Nucl. Phys. B {\bf
	537}, 443 (1999). 
\bibitem{Shuster}
	E.~Shuster and D.T.~Son, Nucl. Phys. B {\bf573}, 434 (2000);
	B.~Park, M.~Rho, A.~Wirzba and I.~Zahed, Phys. Rev. D {\bf 62}, 
	034015 (2000).
\bibitem{Schafer}
	 T.~Sch\"afer, Nucl. Phys. B {\bf 575}, 269 (2000).
\bibitem{Hsu1}
	N.~Evans, J.~Hormuzdiar, S.D.H.~Hsu and M.~Schwetz, 
	Nucl. Phys. B {\bf 581}, 391 (2000). 
\bibitem{Hsu2}
	S.D.H.~Hsu, F.~Sannino and M.~Schwetz,
	Mod. Phys. Lett. A {\bf 16} 1871 (2001).
\bibitem{Hong} 
	D-K.~Hong and S.D.H.~Hsu, Phys. Rev. D {\bf 66}, 071501 (2002);
	D-K.~Hong and S.D.H.~Hsu, Phys. Rev. D {\bf 68}, 034011 (2003).
\bibitem{unlock_SW}
	T.~Sch\"afer and F.~Wilczek, Phys. Rev. D {\bf 60}, 074014 (1999).
\bibitem{unlock_AR}
	M.~Alford, J.~Berges and K.~Rajagopal, Nucl. Phys. B {\bf 558}, 219
	(1999).
\bibitem{Kcond}
	T.~Sch\"afer, Phys. Rev. Lett. {\bf 85}, 5531 (2000);
\bibitem{LOFF}	
	M.G.~Alford, J.A.~Bowers and K.~Rajagopal, Phys. Rev. D {\bf
	  63}, 074016 (2001).
\bibitem{g2SC}
	I.~Shovkovy and M.~Huang, Phys. Lett. B {\bf 564}, 205 (2003);
	M.~Huang and I.~Shovkovy, arXiv:hep-ph/0307273.
\bibitem{gCFL}
	E.~Gubankova, W.V.~Liu and F.~Wilczek, Phys. Rev. Lett. {\bf
	91}, 032001 (2003); 
	M.~Alford, C.~Kouvaris and K.~Rajagopal, arXiv:hep-ph/0311286.
\bibitem{Matsuura} K.~Iida, T.~Matsuura, M.~Tachibana and T.~Hatsuda, 
	arXiv:hep-ph/0312363.
\bibitem{AbukiHatsuda} 
	M.~Matsuzaki, Phys. Rev. D {\bf 62}, 017501 (2000);
	H.~Abuki, T.~Hatsuda and K.~Itakura,
	Phys. Rev. D {\bf 65}, 074014 (2002).
\bibitem{HabukiSD}H.~Abuki, Prog. Theor. Phys. {\bf 110}, 937 (2003).
\bibitem{Shov} 
	I.A.~Shovkovy and L.C.R.~Wijewardhana, Phys. Lett. B {\bf
	470}, 189 (1999). 
\bibitem{Gatto}
	R.~Casalbuoni, R.~Gatto, G.~Nardulli and M. Ruggieri, Phys. Rev. D {\bf
	68}, 034024 (2003). 
\bibitem{CN}
	K.~Iida and G.~Baym, Phys. Rev. D {\bf 63}, 074018 (2001).
	M.~Alford and K.~Rajagopal, JHEP {\bf 0206}, 031 (2002);
	A.W.~Steiner, S.~Reddy and M.~Prakash, Phys. Rev. D {\bf 66},
	094007 (2002);
	F.~Neumann, M.~Buballa and M.~Oertel, 
	Nucl. Phys. A {\bf 714}, 481 (2003);
\end{thebibliography}
\end{document}